\documentclass[copyright,creativecommons]{eptcs}
\providecommand{\event}{EXPRESS 2009} 
\providecommand{\volume}{??}
\providecommand{\anno}{2009}
\providecommand{\firstpage}{1}
\providecommand{\eid}{??}

\usepackage{breakurl}        
\usepackage{amssymb,amsmath,latexsym,amsthm}
\usepackage{color}
\usepackage[all,color,line]{xy} 

\newtheorem{theorem}{Theorem}

\newtheorem{definition}[theorem]{Definition}
\newtheorem{lemma}[theorem]{Lemma}

\newcommand{\move}[1]{\stackrel{#1}{\rightarrow}}

\newcommand{\true}{\textrm{{\it tt}}}
\newcommand{\cA}{{\cal A}}
\newcommand{\cB}{{\cal B}}

\newcommand{\size}{\mathit{size}}

\newcommand{\LTL}{{\mathrm{LTL}}}
\newcommand{\LIO}{{\mathrm{LIO}}}
\newcommand{\s}{_\mathsf{s}}    
\newcommand{\X}{\mathsf{X}}     
\newcommand{\U}{{\,\uU\,}}      
\newcommand{\uUp}{{\uU_{\!\mathsf{+}}}}  
\newcommand{\uU}{\mathsf{U}}    
\newcommand{\F}{\mathsf{F}}     
\newcommand{\G}{\mathsf{G}}     
\newcommand{\Gs}{{\G\s}}        

\newcommand{\bigO}{{\cal O}}
\newcommand{\PSPACE}{\mathsf{PSPACE}}

\newcommand{\coNP}{\mathsf{co{-}NP}}

\newcommand{\AP}{\mathit{A\hskip-0.1ex P}}

\newcommand{\dLTL}{\LTL^\textrm{det}}

\newcommand{\urltilda}{\kern -.15em\lower .7ex\hbox{\~{}}\kern .04em}

\title{Almost Linear B\"uchi Automata} 

\author{Tom\'{a}\v{s} Babiak\thanks{The co-author has been partially
    supported by the Ministry of Education of the Czech Republic, project
    No.~MSM0021622419.}\qquad\qquad Vojt\v{e}ch~{\v{R}}eh\'{a}k\thanks{The
    co-author has been partially supported by the research centre
    ``Institute for Theoretical Computer Science (ITI)'', project
    No.~1M0545, and by the Czech Science Foundation, grant No.~201/08/P459.}
  \qquad\qquad Jan Strej\v{c}ek\thanks{The co-author has been partially
    supported by the Ministry of Education of the Czech Republic, project
    No.~MSM0021622419, and by the Czech Science Foundation, grant
    No.~201/08/P375.}  \institute{Faculty of Informatics\\ Masaryk
    University\\ Brno, Czech Republic}
  \email{\{xbabiak,\,rehak,\,strejcek\}@fi.muni.cz} }

\def\titlerunning{Almost Linear B\"uchi Automata}
\def\authorrunning{T.~Babiak \& V.~\v{R}eh\'{a}k \& J.~Strej\v{c}ek}

\begin{document}
\maketitle
\sloppy


\begin{abstract}
  We introduce a new fragment of Linear temporal logic (LTL) called
  \emph{LIO} and a new class of B\"uchi automata (BA) called \emph{Almost
    linear B\"uchi automata} (ALBA). We provide effective translations
  between LIO and ALBA showing that the two formalisms are expressively
  equivalent. While standard translations of LTL into BA use some
  intermediate formalisms, the presented translation of LIO into ALBA is
  direct. As we expect applications of ALBA in model checking, we compare
  the expressiveness of ALBA with other classes of B\"uchi automata studied
  in this context and we indicate possible applications.
\end{abstract}



\section{Introduction}

The growing number of concurrent software and/or hardware systems puts more
emphasis on development of automatic verification methods applicable in
practice. One of the most promising methods is LTL model checking. The main
problem of this verification method is the \emph{state explosion problem}
and consequent high computational complexity. While symbolic approaches to
model checking partly solve the problem for hardware systems, there is still
no satisfactory solution for model checking of software systems. The most
promising approach seems to be a combination of abstraction methods,
reduction methods, and optimized model checking algorithms. Reduction
methods and optimizations of the algorithms are often based on some specific
properties of the specification formula or the model. For example, one of
the most effective reduction methods called \emph{partial order reduction}
employs the fact that specification formulae usually do not use the modality
\emph{next} and thus they describe \emph{stutter-invariant}
properties~\cite{Lam83}.


We have realized that all formulae of the \emph{restricted temporal
  logic}~\cite{PP04} (i.e.~formulae using only temporal operators
\emph{eventually} and \emph{always}) can be translated to B\"uchi automata
that are linear (1-weak), possibly with an exception of terminal strongly
connected components. These terminal components have also a specific
property: they accept only infinite words over a set of letters, where some
selected letters appear infinitely often. We call such automata \emph{Almost
  linear B\"uchi automata (ALBA)}. In this paper we study mainly the
expressive power of these automata. 

Searching for the precise class of LTL formulae corresponding to ALBA
automata results in the definition of an LTL fragment named \emph{LIO} (the
abbreviation for \emph{linear} and \emph{infinitely often}). The fragment is
strictly more expressive than the restricted temporal logic. To prove that
LIO corresponds to ALBA, we present translations between LIO and ALBA. While
standard translations of LTL formulae into BA use either generalized B\"uchi
automata~\cite{GPVW95} or alternating 1-weak B\"uchi
automata~\cite{MSS88,Var95} as an intermediate formalism, the presented
translation of LIO to ALBA works directly.
Further, there exist LIO formulae such that the corresponding
B\"uchi automata created by the mentioned standard translations are not
ALBA. 

\begin{paragraph}{Related work}
  Some observations regarding specific structure of B\"uchi automata
  corresponding to some LTL fragments have been already published
  in~\cite{CP03}. The paper states that two classes of Manna and Pnueli's
  hierarchy of temporal properties~\cite{MP90}, namely \emph{guarantee} and
  \emph{persistence} formulae, can be translated into \emph{terminal} and
  \emph{weak} automata, respectively. A B\"uchi automaton is
  \emph{terminal}, if every accepting state has a loop transition under each
  letter. An automaton is weak if each strongly connected component consists
  either of accepting or non-accepting states. The paper also suggests some
  improvements of the standard model checking algorithms employing the
  specific structure of the considered property automata. Let us note that
  LIO is incomparable with both guarantee and persistence formulae.
\end{paragraph}

\medskip The paper is structured as follows. Section~\ref{sec:LTL} recalls
the definition of LTL and introduces LIO. Various kinds of B\"uchi automata
including almost linear BA are defined in Section~\ref{sec:BA}. Translations
are presented in Section~\ref{ALBA2LIO} (ALBA~$\rightarrow$~LIO)
and Section~\ref{LIO2ALBA} (LIO~$\rightarrow$~ALBA). Section~\ref{sec:concl} sums up
the presented results and mentions some topics for future research.


\section{Linear temporal logic (LTL)}\label{sec:LTL}

The syntax of \emph{Linear Temporal Logic} (LTL)~\cite{Pnu77} is defined as
follows
\[
  \varphi~::=~\true~\mid~a~\mid~\neg\varphi~\mid~\varphi\vee\varphi~\mid~
  \varphi\wedge\varphi~\mid~\F\varphi~\mid~\G\varphi~\mid~\X\varphi~\mid~
  \varphi\U\varphi\textrm{,}
\]
where $\true$ stands for \emph{true}, $a$ ranges over a countable set $\AP$
of \emph{atomic propositions}, $\F$, $\G$, $\X$, and $\uU$ are modal
operators called \emph{eventually}, \emph{always}, \emph{next}, and
\emph{until}, respectively. The logic is interpreted over infinite words
over the alphabet $\Sigma=2^{\AP'}$, where $\AP'\subseteq\AP$ is a finite
subset. Given a word $u=u(0)u(1)u(2)\ldots\in(2^{\AP'})^\omega$, by $u_i$ we
denote the $i^{th}$ suffix of $u$, i.e.~$u_i=u(i)u(i+1)\ldots$.

The semantics of LTL formulae is defined inductively as follows:
\begin{center}
\begin{tabbing}
  \hspace*{1em} \= $u\models\true$\\
  \> $u\models a$ \hspace*{3em} \= iff \hspace*{.5em} \= $a\in u(0)$\\
  \> $u\models\neg\varphi$ \> iff \> $u\not\models\varphi$\\
  \> $u\models\varphi_1\vee\varphi_2$ \> iff \>
  $u\models\varphi_1$ or $u\models\varphi_2$\\
  \> $u\models\varphi_1\wedge\varphi_2$ \> iff \>
  $u\models\varphi_1$ and $u\models\varphi_2$\\
  \> $u\models\F\varphi$ \> iff \> $\exists i\ge 0\,.\,u_i\models\varphi$\\
  \> $u\models\G\varphi$ \> iff \> $\forall i\ge 0\,.\,u_i\models\varphi$\\
  \> $u\models\X\varphi$ \> iff \> $u_1\models\varphi$\\
  \> $u\models\varphi_1\U\varphi_2$ \> iff \>
     $\exists i\ge 0\,.\,(\,u_i\models\varphi_2$ and
     $\forall\, 0\leq j<i\,.~u_j\models\varphi_1\,)$
\end{tabbing}
\end{center}
We say that a word $u$ \emph{satisfies} $\varphi$ whenever
$u\models\varphi$. Given an alphabet $\Sigma$, a formula $\varphi$ defines
the language $$L^\Sigma(\varphi)=\{u\in\Sigma^\omega\mid u\models\varphi\}.$$

For a set $\{O_1,\ldots,O_n\}$ of modalities, $\LTL(O_1,\ldots,O_n)$ denotes
the LTL fragment containing all formulae with modalities $O_1,\ldots,O_n$
only. We will use mainly the fragments $\LTL(\F,\G)$ with modalities
eventually and always and $\LTL()$ without any modalities. Note that an
$\LTL()$ formula describes only a property of the first letter of an
infinite word. Hence, we say that a letter $e\in\Sigma$ satisfies an
$\LTL()$ formula $\alpha$, written $e\models\alpha$ iff $ew\models\alpha$
for some $w\in\Sigma^\omega$.

\subsection{The LIO fragment}

The \emph{LIO} fragment is defined as
\[
 \varphi~::=~\psi~\mid~\varphi\vee\varphi~\mid~
 \varphi\wedge\varphi~\mid~\X\varphi~\mid~\alpha\U\varphi\textrm{,}
\]
where $\psi$ ranges over $\LTL(\F,\G)$ and $\alpha$ over $\LTL()$.

The fragment does not fit into any standard taxonomy of LTL fragments
(see~\cite{Str04}), but it is a generalization of two standard LTL
fragments:
\begin{itemize}
\item $\LTL(\F,\G)$ - the fragment of all LTL formulae using operators $\F$
  and $\G$ only. This fragment is also known as \emph{restricted temporal
    logic}~\cite{PP04}.
\item flat$\LTL^+(\uU,\X)$ - the fragment of all flat $\LTL(\uU,\X)$
  formulae in positive form. A formula is \emph{flat}~\cite{Dam99} if the
  left subformula of each $\uU$ operator is from $\LTL()$. A formula is in
  \emph{positive} form if there is no modal operator in the scope of any
  negation.
\end{itemize}
In Subsection~\ref{ssec:hierarchy} we show that LIO contains also all
languages expressible as negations of $\LTL^\text{det}$ formulae. The
fragment $\LTL^\text{det}$ is better known as the common fragment of CTL and
LTL~\cite{Mai00}.

The LIO fragment covers many specification formulae frequently used in the
context of model checking, for example typical response formulae of the form
$\G(a\Rightarrow\F b)$. In fact, it is more important that LIO contains
negations of these formulae, as only the negations needs to be translated
into B\"uchi automata.


\section{B\"uchi automata (BA)}\label{sec:BA}

\begin{definition}
  A \emph{B\"uchi automaton} (BA or \emph{automaton} for short) is a
  tuple $A=(\Sigma,Q,q_0,\delta,F)$, where
  \begin{itemize}
  \item $\Sigma$ is a finite \emph{alphabet},
  \item $Q$ is a finite set of \emph{states},
  \item $q_0\in Q$ is an \emph{initial state},
  \item $\delta: Q \times \Sigma \rightarrow 2^Q$ is a \emph{transition
      function}, and
  \item $F \subseteq Q$ is a set of \emph{accepting states}.
  \end{itemize}
\end{definition}
We usually write $p\move{e}q$ instead of $q\in\delta(p,e)$. A B\"uchi
automaton is traditionally seen as a directed graph where nodes are the
states and there is an edge leading from $p$ to $q$ and labelled by $e$
whenever $p\move{e}q$. An edge $p\move{e}p$ is called a \emph{loop} on~$p$.

A \emph{run} $\pi$ over an infinite word
$u(0)u(1)u(2)\ldots\in\Sigma^\omega$ is a sequence
$$\pi=r_0\move{u(0)}r_1\move{u(1)}r_2\move{u(2)}\ldots$$
where $r_0=q_0$ is the initial state. The run is \emph{accepting} if some
accepting state occurs infinitely often in the sequence $r_0,r_1,\ldots$.
The \emph{language} $L(A)$ defined by automaton $A$ is the set of all
infinite words $u$ such that the automaton has an accepting run over $u$.

A state $q$ is \emph{reachable} from $p$, written $p\move{*}q$, if $p=q$ or
there exists a sequence
$$r_0\move{u(0)}r_1\move{u(1)}r_2\move{u(2)}\ldots\move{u(n)}r_{n+1}$$ 
where $p=r_0$ and $q=r_{n+1}$.

A \emph{strongly connected component} (SCC or \emph{component} for short) is
a maximal set of states $S\subseteq Q$ such that $p\move{*}q$ holds for
every $p,q\in S$. Note that every state of an automaton belongs to exactly
one strongly connected component.

Several special classes of B\"uchi automata have been considered in the
context of model checking so far. A B\"uchi automaton
$(\Sigma,Q,q_0,\delta,F)$ is called
\begin{itemize}
\item \emph{terminal} if for each $p\in F$ and $a\in\Sigma$ it holds that
  $\delta(p,a)\neq\emptyset$ and $\delta(p,a)\subseteq F$,
\item \emph{weak} if every SCC of the automaton contains only
  accepting states or only non-accepting states,
\item \emph{$k$-weak} for some $k>0$ if it is weak and every SCC contains at most
  $k$ states,
\item \emph{linear} or \emph{very weak} if it is 1-weak.
\end{itemize}
Linear B\"uchi automata can be alternatively defined as automata where each
SCC consists of one state, i.e.~each cycle is a loop.

Given an automaton $A$ and its state $q$, by $A_q$ we denote the automaton
$A$ where the initial state is changed to $q$. Further, a strongly connected
component $S$ is called \emph{terminal} if for all $p\in S$ it holds that
$p\move{a}q$ implies $q\in S$. To improve the notation, we often label a
transition of a B\"uchi automaton with an $\LTL()$ formula $\alpha$ meaning
that there is a transition under each $e\in\Sigma$ satisfying $\alpha$.

\subsection{Almost linear B\"uchi automata (ALBA)} 

In this section we introduce a new kind of B\"uchi automata and describe its
relation to the previously defined types. 

\begin{definition}
  \emph{Almost linear B\"uchi automaton} (ALBA) is a B\"uchi automaton $A$
  over an alphabet $\Sigma=2^{\AP'}$ 
  such that every non-terminal SCC contains just one state and for every
  terminal component $S$ there exists a formula 
  $$\rho=\G\alpha_0~\wedge\bigwedge_{1\le i\le n}\G\F\alpha_i$$ such that $n\ge 0$,
  $\alpha_0,\alpha_1,\ldots,\alpha_n\in\LTL()$, and for every $q\in S$ it holds
  that $L(A_q)=L^\Sigma(\rho)$.
\end{definition}

Note that our condition on terminal components does not describe their
concrete structure. In fact, a formula $\G\alpha_0\wedge\bigwedge_{0<i\le
  n}\G\F\alpha_i$ can be translated into a (B\"uchi automaton with a single)
component in at least three reasonable ways. We illustrate them by automata
corresponding to the formula $\rho=\G\true\wedge\G\F a_1\,\wedge\,\G\F a_2$.

\begin{figure}[tb]
  \begin{minipage}[b]{0.4\linewidth}
    \[\xymatrix@R=2ex{
      {} \ar[d] \\
      *+=<5ex>[o][F-]{} \ar@(ur,r)^(.7){\neg a_1} \ar[dd]^{a_1}\\ \\
      *+=<5ex>[o][F-]{1} \ar@(ur,r)^(.7){\neg a_2} \ar[dd]^{a_2}\\ \\
      *+=<5ex>[o][F=]{12} \ar@(l,l)[uuuu]^\true 
    }\]
    \vspace{2ex}     
    \caption{Minimal number of transitions.}
    \label{fig:ex1}
  \end{minipage}
  \hfill
  \begin{minipage}[b]{0.5\linewidth}
  \[\xymatrix@R=2ex{
    {} \ar[d] \\
    *+=<5ex>[o][F-]{} \ar@(ur,r)^(.7){\neg a_1} \ar[dd]^{a_1} 
    \ar@/^4pc/[dddd]^{a_1\wedge a_2}\\ \\
    *+=<5ex>[o][F-]{1} \ar@(ur,r)^(.7){\neg a_2} \ar@<.5ex>[dd]^{a_2}\\ \\
    *+=<5ex>[o][F=]{12} \ar@(r,dr)^(.7){a_1\wedge a_2}
    \ar@/^4pc/[uuuu]^{\neg a_1} \ar@<.5ex>[uu]^{a_1\!}
  }\] 
  \caption{Minimal number of states and shortcuts.}
  \label{fig:ex2}
  \end{minipage}
\end{figure}
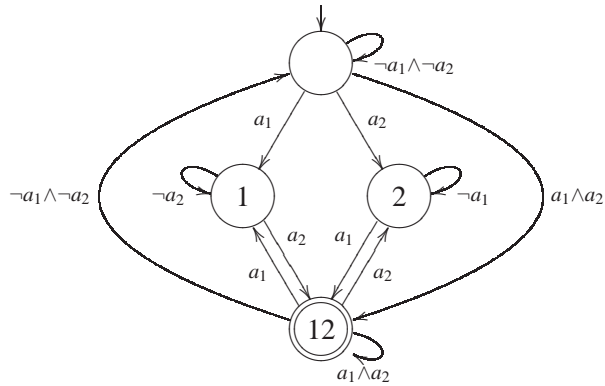
\begin{figure}[htb]
  \[\xymatrix@C=1ex@R=2ex{
    & {} \ar[d] \\
    & *+=<5ex>[o][F-]{} \ar@(ur,r)^(.7){\neg a_1\wedge\neg a_2} \ar[ddl]_{a_1} 
    \ar[ddr]^{a_2} \ar@/^7pc/[dddd]^{a_1\wedge a_2}\\ \\
    *+=<5ex>[o][F-]{1} \ar@(ul,l)_(.7){\neg a_2} \ar@<.5ex>[ddr]^(.4){a_2} && 
    *+=<5ex>[o][F-]{2} \ar@(ur,r)^(.7){\neg a_1} \ar@<-.5ex>[ddl]_(.4){a_1}\\ \\
    & *+=<5ex>[o][F=]{12} \ar@(r,dr)^(.7){a_1\wedge a_2} 
    \ar@/^7pc/[uuuu]^{\neg a_1\wedge\neg a_2} 
    \ar@<.5ex>[uul]^{a_1} \ar@<-.5ex>[uur]_{a_2}
  }\]
  \caption{Shortest cycles in product automata.}
  \label{fig:ex3}  
\end{figure}

\begin{enumerate}
\item If we want to minimize the number of transitions
  and states of the automaton, we create just a ``cycle'' depicted on
  Figure~\ref{fig:ex1}.
\item In the context of LTL model checking, a B\"uchi automaton $A$ derived
  from an LTL formula is usually used to build a \emph{product automaton}
  that accepts all words accepted by $A$ and corresponding to some behaviour
  of the verified system. Model checking algorithms then decide whether
  there is an accepting cycle in the product automaton or not. If we want to
  keep the number of states of $A$ minimal and to shorten the length of
  potential cycles in product automata, we add to the automaton $A$ some
  shortcuts, see Figure~\ref{fig:ex2}.
\item If we want to minimize the length of potential cycles in product
  automata without regard to the number of states, we translate the formula
  $\rho$ into the automaton given in Figure~\ref{fig:ex3}. Note that the
  number of states is exponential in the length of $\rho$, while it is only
  linear in the previous two cases.
\end{enumerate}
In practice, the second kind of translation is usually chosen.

\subsection{Hierarchy of B\"uchi automata classes}\label{ssec:hierarchy}

\begin{figure}[htb]
  \centering
    $\xymatrix@C=20mm@R=6mm{
      & \textrm{(general) BA} \\
      & \textrm{weak BA} \ar@{-}[u] \\
      \textrm{ALBA} \ar@{-}[uur] & \vdots \ar@{-}[u] \\
      & \textrm{3-weak BA} \ar@{-}[u] \\
      & \textrm{2-weak BA} \ar@{-}[u] \\
      & \textrm{linear BA (1-weak BA)} \ar@{-}[u] \ar@{-}[uuul] 
      & \textrm{terminal BA} \ar@{-}[uuuul] 
    }$ 
  \caption{Hierarchy of B\"uchi automata classes.}
  \label{fig:BAhier}
\end{figure}
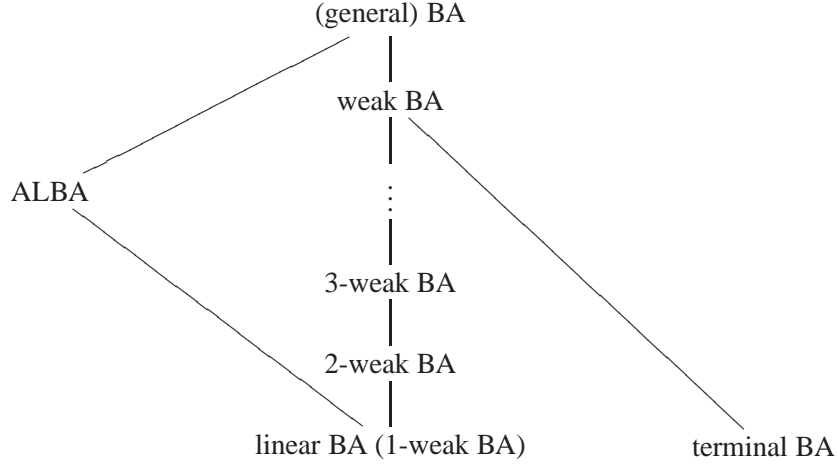
Figure~\ref{fig:BAhier} depicts the hierarchy of the mentioned classes of
B\"uchi automata. A line between two classes means that the upper class is
strictly more expressible than the lower class. If the figure does not
indicate such a relation between a pair of classes, then the classes are
incomparable.

Indicated inclusions follow directly from definitions of the classes. The
strictness of these inclusions is always easy to prove and the same holds
also for the indicated incomparability relations. Note that only two of the
considered classes can express the language of the formula $\G\F a$: ALBA
and the general class.

It is worth mentioning that the class of linear BA is expressively
equivalent to negations of $\LTL^\text{det}$ formulae~\cite{Mai00}.


\section{Translation ALBA~$\rightarrow$~LIO}\label{ALBA2LIO}

Let $A=(\Sigma,Q,q_0,\delta,F)$ be an ALBA. For every state $q\in Q$, we
recursively define a LIO formula $\varphi(q)$ such that
$L(A_q)=L^\Sigma(\varphi(q))$. There are two cases:
\begin{itemize}
\item $q$ is in a terminal strongly connected component. Due to the
  definition of ALBA, there exists a
  formula $$\rho=\G\alpha_0~\wedge\bigwedge_{1\le i\le n}\G\F\alpha_i$$ such that
  $n\ge 0$, $\alpha_0,\alpha_1,\ldots,\alpha_n\in\LTL()$. We set
  $\varphi(q)=\rho$. Note that $\rho$ is a formula of $\LTL(\F,\G)$.
\item $q$ is not in any terminal component. Let $q\move{a_1}q$,
  $q\move{a_2}q$, $\ldots$, $q\move{a_n}q$ be all loops on $q$ and
  $q\move{b_1}q_1$, $q\move{b_2}q_2$, $\ldots$, $q\move{b_m}q_m$ be all
  transitions leading from $q$ to other states. Then we set
  \[
    \varphi(q)=\left\{
      \begin{array}{lp{2ex}l}
        \displaystyle
        (\bigvee_{0<i\le n} a_i)~\U\bigvee_{0<j\le m}(b_j\wedge\X\varphi(q_j))
        && \text{if }q\not\in F\!\text{,} \\ \\
        \displaystyle       
        \Big((\bigvee_{0<i\le n} a_i)~\U\bigvee_{0<j\le m}(b_j\wedge\X\varphi(q_j))\Big)
        ~~~\vee~~~\G\bigvee_{0<i\le n} a_i
        && \text{if }q\in F. 
      \end{array}
    \right.
  \]
  Note that $\varphi(q)$ is in LIO assuming that all $\varphi(q_j)$ are in
  LIO.
\end{itemize}

\medskip
The correctness of the recursion follows from the fact that $A$ is linear
(except the terminal components). The whole automaton then corresponds to
the formula $\varphi(q_0)$.


\section{Translation LIO~$\rightarrow$~ALBA}\label{LIO2ALBA}

In this section, we always assume that LIO formulae are in \emph{positive form},
i.e.~no temporal operator is in scope of any negation. Every LIO formula can
be transformed into this form using the following equivalences.

\bigskip\noindent
$\neg\F\varphi\equiv\G\neg\varphi$ \hfill
$\neg\G\varphi\equiv\F\neg\varphi$ \hfill
$\neg(\varphi_1\wedge\varphi_2)\equiv\neg\varphi_1\,\vee\,\neg\varphi_2$ \hfill
$\neg(\varphi_1\vee\varphi_2)\equiv\neg\varphi_1\,\wedge\,\neg\varphi_2$

\bigskip
For each LIO formula $\varphi$, we define its \emph{size} as follows:
\begin{itemize}
\item if $\varphi$ is in $\LTL()$, we set $\size(\varphi)=1$,
\item if $\varphi$ is not in $\LTL()$, we define its size recursively:
  \[
  \begin{array}{rclr}
    \size(\varphi_1\vee\varphi_2)&=&\size(\varphi_1)+1+\size(\varphi_2) \\
    \size(\varphi_1 \wedge \varphi_2) & = & \size(\varphi_1) + 1 + \size(\varphi_2)\\
    \size(\F\varphi) & = & 1 + \size(\varphi)\\
    \size(\G\varphi) & = & 2*\size(\varphi) \\
    \size(\X\varphi) & = & 1 + \size(\varphi)\\
    \size(\alpha\U\varphi) & = & 1 + \size(\varphi)\\
  \end{array}
  \]
\end{itemize}

Let $S$ be a finite set of LIO formulae. We define
its size as
\[
\begin{array}{rcl}
  \size(\emptyset) & = & (0,-) \\
  \size(S) & = & (k, (i_k,i_{k-1}, \dots , i_1))
\end{array}
\]
where $k = max\{\size(\varphi) \mid \varphi \in S\}$ and $i_j =
\left|\{\varphi \mid \varphi \in S\,\wedge\, \size(\varphi)=j\}\right|$ for
each $k \ge j \ge 1$. Finally, we define a strict (lexicographical) order
$<$ on sizes of these sets in the following way.
\[
\setlength{\arraycolsep}{0pt}
\begin{array}{rcll}
  (k,(i_k,i_{k-1}, \dots , i_1)) < (l,(j_l,j_{l-1}, \dots , j_1))
  &\,\iff\,&\,k<l~\, \vee & \\
  &&\!\!(k=l ~\wedge~ \exists\,k\ge m\ge 1~.~i_m < j_m~\wedge~
  \forall\,k\ge n>m ~.~i_n = j_n)
\end{array}
\]

The translation is based on transformation of a LIO formula into an
equivalent formula of a special form. Formally, to every LIO formula $\varphi$
we assign a set $R(\varphi)\subseteq\LTL()\times P_\mathit{fin}(\LIO)$, where
$P_\mathit{fin}(\LIO)$ is the set of all finite subsets of LIO, such that
$$\varphi~\equiv\bigvee_{(\alpha,S)\in
  R(\varphi)}(\alpha~\wedge~\X\bigwedge_{\sigma\in S}\sigma).$$

 
The set $R(\varphi)$ is defined recursively. The recursion is always bounded
as each $R(\varphi')$ appearing in the definition of $R(\varphi)$ satisfies
$\size(\varphi')<\size(\varphi)$. In the following, $\alpha$ always represents
a formula of $\LTL()$. We define $R(\varphi)$ according to the structure of
$\varphi$.
\begin{itemize}
\item \fbox{$\alpha$}~ $R(\alpha)=\{(\alpha,\emptyset)\}$
\item \fbox{$\varphi_1\vee\varphi_2$}~ $R(\varphi_1\vee\varphi_2)= R(\varphi_1)\cup R(\varphi_2)$
\item \fbox{$\varphi_1\wedge\varphi_2$}~ $R(\varphi_1\wedge\varphi_2)=
    \{(\alpha_1\wedge\alpha_2,S_1\cup S_2)\mid
    (\alpha_1,S_1)\in R(\varphi_1),(\alpha_2,S_2)\in R(\varphi_2)\}$
\item \fbox{$\F\varphi_0$}~ $R(\F\varphi_0)=\{(\true,\{\F\varphi_0\})\}\cup R(\varphi_0)$
\item \fbox{$\X\varphi_0$}~ $R(\X\varphi_0)=\{(\true,\{\varphi_0\})\}$
\item \fbox{$\alpha\U\varphi_0$}~
  $R(\alpha\U\varphi_0)=\{(\alpha,\{\alpha\U\varphi_0\})\}\cup R(\varphi_0)$
\item \fbox{$\G\varphi_0$}~ This case is divided into the following subcases according
  to the structure of $\varphi_0$:
  \begin{itemize}
  \addtolength{\itemsep}{.5ex}
  \item \fbox{$\alpha$}~ $R(\G \alpha)=\{(\alpha,\{\G \alpha\})\}$
  \item \fbox{$\varphi_1\wedge\varphi_2$}~ $R(\G(\varphi_1\wedge\varphi_2))=
    R(\G\varphi_1\,\wedge\,\G\varphi_2)$
  \item \fbox{$\F\varphi_1$}~ This case is again divided into the following
    subcases according to the structure of $\varphi_1$:
    \begin{itemize}
    \item \fbox{$\alpha$}~ $R(\G\F \alpha)=\{(\true,\{\G\F \alpha\})\}$ \strut
    \item \fbox{$\varphi_3\vee\varphi_4$}~ $R(\G\F(\varphi_3\vee\varphi_4))=
      R(\G\F\varphi_3)\cup R(\G\F\varphi_4)$
    \item \fbox{$\varphi_3\wedge\varphi_4$}~ As conjunction is an
      associative operator, we can see it as an operator of arbitrary arity
      and we can assume that all conjuncts are not conjunctions. Then either
      all conjuncts are formulae of $\LTL()$
      (i.e.~$\varphi_3\wedge\varphi_4\in\LTL()$ - this case has been already
      covered by the Case~$\G\F \alpha$\hspace{.1ex}), or at least one
      conjunct has the form $\varphi_5\vee\varphi_6$ or $\F\varphi_5$ or
      $\G\varphi_5$.  Let $\varphi_4$ be this conjunct and $\varphi_3$ be
      conjunction of all the other conjuncts. We proceed according to the
      structure of $\varphi_4$.
      \begin{itemize}
      \item \fbox{$\varphi_5\vee\varphi_6$}~ As
        $\G\F(\varphi_3\wedge(\varphi_5\vee\varphi_6))\equiv
        \G\F(\varphi_3\wedge\varphi_5)\vee \G\F(\varphi_3\wedge\varphi_6)$, we set \\
        $R(\G\F(\varphi_3\wedge(\varphi_5\vee\varphi_6)))=
        R(\G\F(\varphi_3\wedge\varphi_5))\cup
        R(\G\F(\varphi_3\wedge\varphi_6))$.
      \item \fbox{$\F\varphi_5$}~ As
        $\G\F(\varphi_3\wedge\F\varphi_5)\equiv(\G\F\varphi_3)\wedge(\G\F\varphi_5)$,
        we set\\
        $R(\G\F(\varphi_3\wedge\F\varphi_5))=R((\G\F\varphi_3)\wedge(\G\F\varphi_5))$.
      \item \fbox{$\G\varphi_5$}~ As $\G\F(\varphi_3\wedge\G\varphi_5)\equiv
        (\G\F\varphi_3)\wedge(\G\F\G\varphi_5)\equiv
        (\G\F\varphi_3)\wedge(\F\G\varphi_5)$, we set \\
        $R(\G\F(\varphi_3\wedge\G\varphi_5))=R((\G\F\varphi_3)\wedge(\F\G\varphi_5))$.
      \end{itemize}
    \item \fbox{$\F\varphi_3$}~ $R(\G\F\F\varphi_3)= R(\G\F\varphi_3)$
    \item \fbox{$\G\varphi_3$}~ $R(\G\F\G\varphi_3)= R(\F\G\varphi_3)$
    \end{itemize}
  \item \fbox{$\varphi_1\vee\varphi_2$}~ The situation is similar to the Case
    $\G\F(\varphi_2\wedge\varphi_4)$. Hence, either
    $\varphi_1\vee\varphi_2\in\LTL()$ (this has been already solved in
    Case~$\G\alpha$), or we can assume that $\varphi_2$ has the form
    $\varphi_3\wedge\varphi_4$ or $\F\varphi_3$ or $\G\varphi_3$.
    We proceed according to the structure of $\varphi_2$.
    \begin{itemize}
    \item \fbox{$\varphi_3\wedge\varphi_4$}~ As $\G(\varphi_1\vee(\varphi_3\wedge\varphi_4))\equiv
      \G(\varphi_1\vee\varphi_3)\wedge\G(\varphi_1\vee\varphi_4)$, we set 
      $R(\G(\varphi_1\vee(\varphi_3\wedge\varphi_4)))=
      \{(\alpha_1\wedge\alpha_2,S_1\cup S_2)\mid(\alpha_1,S_1)\in
      R(\G(\varphi_1\vee \varphi_3)),(\alpha_2,S_2)\in
      R(\G(\varphi_1\vee\varphi_4))\}$.
    \item \fbox{$\F\varphi_3$}~ As $\G(\varphi_1\vee\F\varphi_3)\equiv (\G\varphi_1)\vee
      \F(\varphi_3\wedge\X\G\varphi_1)\vee\G\F\varphi_3 \equiv (\G\varphi_1)\vee
      \true\U(\varphi_3\wedge(\X\G\varphi_1))\vee \G\F\varphi_3$, we set 
      $R(\G(\varphi_1\vee\F\varphi_3))=R(\G\varphi_1)\cup
      R(\true\U(\varphi_3\wedge(\X\G\varphi_1)))\cup R(\G\F\varphi_3)$.
    \item \fbox{$\G\varphi_3$}~ $R(\G(\varphi_1\vee\G\varphi_3))$: Here we
      consider only the following two structures of the whole subformula
      $\varphi_1\vee\G\varphi_3$ (the other possibilities fit to some of the
      previous cases):
      \begin{itemize}
      \item \fbox{$\bigvee_{\varphi'\in G}\G\varphi'$}~ As
        $\G(\bigvee_{\varphi'\in G}\G\varphi')\equiv\bigvee_{\varphi'\in
          G}(\G\varphi')$, we set \\ $R(\G(\bigvee_{\varphi'\in
          G}\G\varphi')) = \bigcup_{\varphi'\in G} R(\G\varphi')$.
      \item \fbox{$\alpha\vee\bigvee_{\varphi'\in G}\G\varphi'$}~ As 
        $\G(\alpha\vee\bigvee_{\varphi'\in G}\G\varphi')\equiv$ \\
        $(\G\alpha)\vee\bigvee_{\varphi'\in G}(\G\varphi')\vee
        \bigvee_{\varphi'\in G}(\alpha\wedge\X\G(\alpha\vee\G\varphi')$, we set \\
        $R(\G(\alpha\vee\bigvee_{\varphi'\in G}\G\varphi'))= R(\G\alpha)
        \cup \bigcup_{\varphi'\in G}R(\G\varphi')\cup \bigcup_{\varphi'\in
          G}\{(\alpha,\{\G(\alpha\vee\G\varphi')\})\}$.
      \end{itemize}
    \end{itemize}
  \item \fbox{$\G\varphi_1$}~ $R(\G\G\varphi_1)=R(\G\varphi_1)$
  \end{itemize}
\end{itemize}
Moreover, for every finite set of LIO formulae, we define 
$$R(S)=R(\bigwedge_{\varphi\in S} \varphi).$$ 
In particular, $R(\emptyset)=\{(\true,\emptyset)\}$.

Before we provide the construction of ALBA automaton for a given LIO
formula, we mention some crucial observations. First of all, one can readily
confirm the following observation.
\begin{lemma}\label{lem:1}
  For every $(\alpha,S)\in R(\varphi)$ it holds that, for each $\varphi'\in
  S$, either $\varphi'=\varphi$ or $\size(\varphi')\,{<}\,\size(\varphi)$.
\end{lemma}
In fact, there are only five cases where $\varphi'=\varphi$, namely if
$\varphi$ has the form $\F\varphi_0$ or $\alpha\U\varphi_0$ or $\G\alpha$ or
$\G\F\alpha$ or $\G(\alpha\vee\G\varphi')$ (this is a special case of the
form $\G(\alpha\vee\bigvee_{\varphi'\in G}\G\varphi')$). Lemma~\ref{lem:1}
together with an analysis of the listed cases directly implies the following
property.
\begin{lemma}\label{lem:2}
  For every $(\alpha,S)\in R(\{\varphi\})$, either $S=\{\varphi\}$ or
  $\size(S)<\size(\{\varphi\})$.
\end{lemma}
This lemma immediately implies the following one.
\begin{lemma}\label{lem:3}
  Let $S$ be a finite set of LIO formulae in positive form. For every
  $(\alpha,S')\in R(S)$ it holds that $S=S'$ or $\size(S')<\size(S)$.
\end{lemma}
If we look at the five cases mentioned above Lemma~\ref{lem:2}, we can
easily see that only two of them have the property that
$R(\varphi)=(\alpha,\varphi)$, namely the cases $\G\alpha$ and
$\G\F\alpha$. This is the crucial argument for the following observation.
\begin{lemma}\label{lem:4}
  Let $S$ be a finite set of LIO formulae in positive form. It holds that 
  $$S\subseteq\{\G\alpha,\G\F\alpha\mid\alpha\in
  \LTL()\}~~~\textrm{iff}~~~(\alpha,S')\in R(S) \,\Rightarrow\, S=S'.$$
\end{lemma}

\medskip Now we are ready to finish the translation. Let $\varphi$ be a LIO
formula in positive normal form and let $\AP'$ be the set of all atomic
propositions occurring in $\varphi$. We describe the ALBA automaton in a
concise form: terminal components will be described by distinguished states
labelled with the corresponding LTL formulae of the form
$\rho=\G\alpha_0~\wedge\bigwedge_{1\le i\le n}\G\F\alpha_i$. A standard ALBA
can be obtained from this concise form very easily: we just replace every
such a state by a corresponding component (as indicated in
Figures~\ref{fig:ex1}, \ref{fig:ex2}, or~\ref{fig:ex3}).

The automaton corresponding to $\varphi$ is constructed as
$(\Sigma,Q,q_0,\delta,F)$, where
\begin{itemize}
\item $\Sigma=2^{\AP'}$,
\item $Q=2^M$ and $M=\{\varphi'\mid \varphi'~\textrm{is a LIO formula
    over}~\AP'~\textrm{and}~\size(\varphi')\le\size(\varphi)\}$ is a set
  subsuming all formulae that can be derived from $\varphi$ by repeated
  applications of $R(\cdot)$ (see Lemma~\ref{lem:1}),
\item $q_0=\{\varphi\}$,
\item For each $e\in\Sigma$ and $S\in Q$, we set 
  $\delta(S,e)=\{S'\mid (\alpha,S')\in R(S)~\textrm{and}~e\models\alpha\}$, 
\item accepting states appear only in terminal components. Due to
  Lemma~\ref{lem:4}, terminal components correspond to states $S$ satisfying
  $S\subseteq\{\G\alpha,\G\F\alpha\mid\alpha\in\LTL()\}$. Hence, we label
  such a state $S$ with the formula $$(\G\bigwedge_{\G\alpha\in
    S}\alpha)\wedge\bigwedge_{\G\F\alpha\in S}\G\F\alpha$$ of the desired
  form.
\end{itemize}
The language equivalence between $\varphi$ and the constructed automaton
follows from the properties of $R(\cdot)$. The constructed automaton is ALBA
due to Lemma~\ref{lem:3} (linearity except terminal components) and
Lemma~\ref{lem:4} (condition on terminal components). Note that the
translation directly provides triple exponential bound on the size of $Q$ in
the length of $\varphi$ (even $\size(\varphi)$ can be exponential in the
length of $\varphi$). However, we conjecture that the size of $Q$ is in fact
only singly exponential in the length of $\varphi$.


\medskip A natural question is whether standard translations of LTL into BA
also produce ALBA when applied to LIO. The answer is negative. For example,
Gastin and Oddoux's popular implementation of the translation via
alternating automata~\cite{GO01} (available online at
\texttt{http://www.lsv.ens-cachan.fr/\urltilda gastin/ltl2ba/index.php})
transforms the LIO formula $\G(\G(a\vee \F b)\vee \G(c\vee\F d))$ into a
B\"uchi automaton that is not ALBA (it contains nonterminal strongly
connected components of size greater than one).


\section{Conclusion}\label{sec:concl}

We have introduced a new class of B\"uchi automata called \emph{almost
  linear B\"uchi automata (ALBA)}. We have compared the expressive power of
ALBA with other classes of B\"uchi automata. Further, we have identified a
fragment of LTL called \emph{LIO} and equivalent to ALBA. The LIO fragment
subsumes some previously studied LTL fragments, in particular the
\emph{restricted temporal logic} and negations of $\dLTL$, i.e.~the common
fragment of CTL and LTL. We have provided a direct translation of LIO
formulae into B\"uchi automata (BA). In contrast to standard translations of
LTL into BA, our translation does not use any intermediate formalism and
always produces ALBA. We expect that the specific structure of ALBA can lead
to development of algorithms designed especially for model checking of
negations of LIO properties. To emphasize potential usability of such
algorithms, we have analysed the collection of the most often verified
properties called Specification Patterns~\cite{DAC98}. It shows up that
negations of 89\% of the properties can be expressed as LIO formulae and
hence translated to ALBA.




\bibliographystyle{eptcs} 
\bibliography{icalp}

\begin{thebibliography}{10}
\providecommand{\bibitemstart}[1]{\bibitem{#1}}
\providecommand{\bibitemend}{}
\providecommand{\bibliographystart}{}
\providecommand{\bibliographyend}{}
\providecommand{\url}[1]{\texttt{#1}}
\providecommand{\urlprefix}{Available at }
\providecommand{\bibinfo}[2]{#2}
\bibliographystart

\bibitemstart{CP03}
\bibinfo{author}{Ivana {\v{C}}ern\'a} \& \bibinfo{author}{Radek Pel\'anek}
  (\bibinfo{year}{2003}): \emph{\bibinfo{title}{Relating Hierarchy of Temporal
  Properties to Model Checking}}.
\newblock In: {\sl \bibinfo{booktitle}{Proceedings of the 30th Symposium on
  Mathematical Foundations of Computer Science ({MFCS}'03)}}, {\sl
  \bibinfo{series}{Lecture Notes in Computer Science}} \bibinfo{volume}{2747}.
  \bibinfo{publisher}{Springer-Verlag}, pp. \bibinfo{pages}{318--327}.
\bibitemend

\bibitemstart{Dam99}
\bibinfo{author}{Dennis~R. Dams} (\bibinfo{year}{1999}):
  \emph{\bibinfo{title}{Flat Fragments of {CTL} and {CTL}$^*$: Separating the
  Expressive and Distinguishing Powers}}.
\newblock {\sl \bibinfo{journal}{Logic Journal of the IGPL}}
  \bibinfo{volume}{7}(\bibinfo{number}{1}), pp. \bibinfo{pages}{55--78}.
\bibitemend

\bibitemstart{DAC98}
\bibinfo{author}{Matthew~B. Dwyer}, \bibinfo{author}{George~S. Avrunin} \&
  \bibinfo{author}{James~C. Corbett} (\bibinfo{year}{1998}):
  \emph{\bibinfo{title}{Property Specification Patterns for Finite-State
  Verification}}.
\newblock In: {\sl \bibinfo{booktitle}{Proc.~2nd Workshop on Formal Methods in
  Software Practice ({FMSP}-98)}}. \bibinfo{publisher}{ACM Press},
  \bibinfo{address}{New York}, pp. \bibinfo{pages}{7--15}.
\bibitemend

\bibitemstart{GO01}
\bibinfo{author}{Paul Gastin} \& \bibinfo{author}{Denis Oddoux}
  (\bibinfo{year}{2001}): \emph{\bibinfo{title}{Fast {LTL} to {B}{\"u}chi
  Automata Translation}}.
\newblock In: \bibinfo{editor}{G.~Berry}, \bibinfo{editor}{H.~Comon} \&
  \bibinfo{editor}{A.~Finkel}, editors: {\sl \bibinfo{booktitle}{Proceedings of
  the 13th International Conference on Computer Aided Verification
  ({CAV}'01)}}, {\sl \bibinfo{series}{Lecture Notes in Computer Science}}
  \bibinfo{volume}{2102}. \bibinfo{publisher}{Springer-Verlag}, pp.
  \bibinfo{pages}{53--65}.
\bibitemend

\bibitemstart{GPVW95}
\bibinfo{author}{Rob Gerth}, \bibinfo{author}{Doron Peled},
  \bibinfo{author}{Moshe~Y. Vardi} \& \bibinfo{author}{Pierre Wolper}
  (\bibinfo{year}{1995}): \emph{\bibinfo{title}{Simple On-the-fly Automatic
  Verification of Linear Temporal Logic}}.
\newblock In: {\sl \bibinfo{booktitle}{Protocol Specification Testing and
  Verification}}. \bibinfo{publisher}{Chapman \& Hall}, pp.
  \bibinfo{pages}{3--18}.
\bibitemend

\bibitemstart{Lam83}
\bibinfo{author}{Leslie Lamport} (\bibinfo{year}{1983}):
  \emph{\bibinfo{title}{What good is Temporal Logic?}}
\newblock In: \bibinfo{editor}{{R. E. A. Mason}}, editor: {\sl
  \bibinfo{booktitle}{Proceedings of the {IFIP} Congress on Information
  Processing}}. \bibinfo{publisher}{North-Holland},
  \bibinfo{address}{Amsterdam}, pp. \bibinfo{pages}{657--667}.
\bibitemend

\bibitemstart{Mai00}
\bibinfo{author}{Monika Maidl} (\bibinfo{year}{2000}):
  \emph{\bibinfo{title}{The common fragment of {CTL} and {LTL}}}.
\newblock In: \bibinfo{editor}{D.~C. Young}, editor: {\sl
  \bibinfo{booktitle}{Proceedings of the 41st Annual {IEEE} Symposium on
  Foundations of Computer Science ({FOCS}'00)}}. \bibinfo{publisher}{IEEE
  Computer Society Press}, pp. \bibinfo{pages}{643--652}.
\bibitemend

\bibitemstart{MP90}
\bibinfo{author}{Zohar Manna} \& \bibinfo{author}{Amir Pnueli}
  (\bibinfo{year}{1990}): \emph{\bibinfo{title}{A hierarchy of temporal
  properties}}.
\newblock In: {\sl \bibinfo{booktitle}{Proceedings of {ACM} Symposium on
  Principles of Distributed Computing ({PODC}'90)}}. \bibinfo{publisher}{ACM
  Press}, pp. \bibinfo{pages}{377--410}.
\bibitemend

\bibitemstart{MSS88}
\bibinfo{author}{David~E. Muller}, \bibinfo{author}{Ahmed Saoudi} \&
  \bibinfo{author}{Paul~E. Schupp} (\bibinfo{year}{1988}):
  \emph{\bibinfo{title}{Weak alternating automata give a simple explanation of
  why most temporal and dynamic logics are decidable in exponential time}}.
\newblock In: {\sl \bibinfo{booktitle}{Proceedings of the 3rd Annual {IEEE}
  Symposium on Logic in Computer Science ({LICS}'88)}}.
  \bibinfo{publisher}{IEEE Computer Society Press}, pp.
  \bibinfo{pages}{422--427}.
\bibitemend

\bibitemstart{PP04}
\bibinfo{author}{Dominique Perrin} \& \bibinfo{author}{Jean-Eric Pin}
  (\bibinfo{year}{2004}): \emph{\bibinfo{title}{Infinite words}}, {\sl
  \bibinfo{series}{Pure and Applied Mathematics}} \bibinfo{volume}{141}.
\newblock \bibinfo{publisher}{Elsevier}.
\bibitemend

\bibitemstart{Pnu77}
\bibinfo{author}{Amir Pnueli} (\bibinfo{year}{1977}): \emph{\bibinfo{title}{The
  temporal logic of programs}}.
\newblock In: {\sl \bibinfo{booktitle}{Proc.~18th IEEE Symposium on the
  Foundations of Computer Science}}. pp. \bibinfo{pages}{46--57}.
\bibitemend

\bibitemstart{Str04}
\bibinfo{author}{Jan Strej\v{c}ek} (\bibinfo{year}{2004}):
  \emph{\bibinfo{title}{Linear Temporal Logic: Expressiveness and Model
  Checking}}.
\newblock \bibinfo{type}{Ph.D. thesis}, \bibinfo{school}{Faculty of
  Informatics, Masaryk University in Brno}.
\bibitemend

\bibitemstart{Var95}
\bibinfo{author}{Moshe~Y. Vardi} (\bibinfo{year}{1995}):
  \emph{\bibinfo{title}{An Automata-Theoretic Approach to Linear Temporal
  Logic}}.
\newblock In: {\sl \bibinfo{booktitle}{Banff Higher Order Workshop}}, {\sl
  \bibinfo{series}{Lecture Notes in Computer Science}} \bibinfo{volume}{1043}.
  \bibinfo{publisher}{Springer}, pp. \bibinfo{pages}{238--266}.
\bibitemend

\bibliographyend
\end{thebibliography}

\end{document}